\documentclass[12pt,preprint]{aastex}

\usepackage{longtable}
\usepackage{amsfonts}
\usepackage{rotating}
\RequirePackage{amssymb}
\RequirePackage{latexsym}
\RequirePackage{graphicx}

\bibpunct{(}{)}{;}{a}{}{,}
\def\vmy{\hbox{\it v--y\/}}
\def\umy{\hbox{\it u--y\/}}
\def\camy{\hbox{\it Ca--y\/}}
\def\bmy{\hbox{\it b--y\/}}
\def\feh{\hbox{\rm [Fe/H]}}
\def\mh{\hbox{\rm [M/H]}}
\def\zx{\hbox{\rm Z/X}}

\def\afe{\hbox{\rm [$\alpha$/Fe]}}
\usepackage{natbib}

\renewcommand{\min}{\mbox{$^m$}}

\def\min{${}^{\prime}$}

\newcommand{\strom}{\mbox{Str\"omgren~}}

\shorttitle{On a new theoretical calibration of the \strom $hk$ metallicity index} 
\shortauthors{Calamida et al.}

\begin{document}



\title{On a new theoretical calibration of the St\"omgren $hk$ metallicity index: 
NGC~6522 as a first test case  
\altaffilmark{1}}

\author{
A. Calamida\altaffilmark{2},
G. Bono\altaffilmark{3}, 
C. E. Corsi\altaffilmark{2},
G. Iannicola\altaffilmark{2},
V. Ripepi\altaffilmark{4},
B. Anthony-Twarog\altaffilmark{5},
B. Twarog\altaffilmark{5},
M. Zoccali\altaffilmark{6},
R. Buonanno\altaffilmark{3},
S. Cassisi\altaffilmark{7},
I. Ferraro\altaffilmark{2}, 
F. Grundahl\altaffilmark{8},
A. Pietrinferni\altaffilmark{7},
L. Pulone\altaffilmark{2}
}

\altaffiltext{1}{Based on observations collected with the 1.54m Danish telescope (ESO, La Silla). 
Period 65}

\altaffiltext{2}{INAF-Osservatorio Astronomico di Roma, Via Frascati 33, 00040, Monte Porzio 
Catone, Italy; annalisa.calamida@oa-roma.inaf.it, corsi@oa-roma.inaf.it, 
ferraro@oa-roma.inaf.it, giacinto@oa-roma.inaf.it, pulone@oa-roma.inaf.it}

\altaffiltext{3}{Universit\`a di Roma Tor Vergata, Via della Ricerca Scientifica 1,
00133 Rome, Italy; bono@roma2.infn.it, buonanno@roma2.infn.it}

\altaffiltext{4}{INAF - Osservatorio Astronomico di Capodimonte,
Via Moiariello 16, 80131 Napoli; ripepi@na.astro.it}

\altaffiltext{5}{Department of Physics and Astronomy, University of Kansas,
1082 Malott,1251 Wescoe Hall Dr., Lawrence, KS 66045-7582; 
bjat@ku.edu, btwarog@ku.edu}

\altaffiltext{6}{Pontificia Universidad Cat\'olica de Chile, 
Departemento de Astronomia y Astrofisica, Avda. Libertador Bernardo OHiggins 340, 
Santiago, Chile; mzoccali@astro.puc.cl}

\altaffiltext{7}{INAF-Osservatorio Astronomico di Collurania, via M. Maggini, 
64100 Teramo, Italy; cassisi@oa-teramo.inaf.it, adriano@oa-teramo.inaf.it}

\altaffiltext{8}{Department of Physics and Astronomy, Aarhus University,
Ny Munkegade, 8000 Aarhus C, Denmark; fgj@phys.au.dk }

\date{\centering ApJL, accepted on \today}

\begin{abstract}
We present a new theoretical calibration of the \strom metallicity index $hk$ 
using $\alpha$-enhanced evolutionary models transformed into the observational 
plane by using atmosphere models with the same chemical mixture. 
We apply the new Metallicity--Index--Color (MIC) relations to a sample of 
85 field red giants (RGs) and find that the difference between photometric 
estimates and spectroscopic measurements is on average smaller than 
0.1 dex with a dispersion of $\sigma$= 0.19 dex.  
The outcome is the same if we apply the MIC relations to a sample 
of eight RGs in the bulge globular cluster NGC~6522, but the standard 
deviation ranges from 0.26 ($hk, \, \vmy$) to 0.49 ($hk, \, \umy$). 
The difference is mainly caused by a difference in photometric accuracy.  
The new MIC relations based on the $\camy$ color provide metallicities 
systematically more metal-rich than the spectroscopic ones.
We found that the $Ca$-band is affected by $Ca$ abundance and 
possibly by chromospheric activity. 
\end{abstract}



\section{Introduction}\label{introduction}
The intermediate-band \strom photometric system \citep{strom66}
has, for stars with spectral types from A to G, several indisputable 
advantages when compared with broad-band photometric systems.

{\bf i\/}) the Str\"omgren index $m_1=(v-b)-(b-y)$ was specifically devised 
to estimate stellar metallicity of both evolved (horizontal branch [HB], 
red-giant branch [RGB]) and main sequence stars \citep[hereafter CA07]{hilkri00,twa00,io07}, 
while the $hk$ index, defined as $(Ca-b)-(b-y)$ \citep{ttwa91},
replaces the $v$ with the $Ca$ filter centered on the $Ca II$ $H$ and $K$ 
lines, and it is {\it primarily} sensitive to the $Ca$ star abundance.
The $hk,\,\bmy$ plane has been adopted to estimate the metallicity of
both RG \citep{twa95} and $RR$ Lyrae stars \citep{baird,rey}.
The main advantage of \strom metallicity indices over stellar spectroscopy 
is that they provide simultaneous metallicity estimates for large samples 
of stars. However, the use of the \strom indices does require precise 
multiband photometry and absolute calibration.    

{\bf ii)} The $v$ filter is strongly affected by two $CN$ molecular absorption 
bands ($\lambda=4142$ and $\lambda=4215$ \AA). Stars with an over-abundance 
of carbon ($C$) and/or nitrogen ($N$), i.e. $CH$- and/or $CN$-strong stars, 
will have, at fixed color, larger $m_1$, a fundamental property for 
identifying stars with different $CNO$ abundances in 
Globular Clusters (GCs, CA07, \citealt{io09}).

{\bf iii)} The $hk$ index is based on the $CaII$ $H,K$ lines that, at fixed
metal abundance, are stronger than weak metallic lines falling across the 
$v$ filter. This means that the $hk$ index in the metal-poor regime is more 
sensitive to metallicity changes than $m_1$. For cool stars, saturation of the
$CaII$ lines leads to a reversal of this trend in the metal-rich regime \citep{twa95}.

{\bf iv)} \citet{lee09b}, using $u,v,b,y, Ca$ data for 37 GGCs, found that 
most of them show a discrete or broad RGB in the $y,\, hk$ plane, and
suggested a spread in $Ca$ and/or heavy element abundance in these GGCs.

\section{Observations and calibration of the $hk$ metallicity index.}\label{observations}

{\it Ca-uvby\/} \strom images were collected during two 
observing nights (July 6-7, $2000$) with the 1.54m Danish Telescope 
(ESO) and the DFOSC camera, with a pixel scale of 0.39\arcsec 
and a field of view of 13.7\min$\times$13.7\min. The pointing was centered 
on the Baade's Window ($\alpha$ = 18:03:34, $\delta$ = -30:04:10), 
including NGC~6522. We secured 16 images ($4y,4b,2v,2Ca,2u$),   
with exposure times ranging from 60s ($y$) to 1000s ($Ca$), 
and seeing between $\sim$1\farcs2 and $\sim$1\farcs6. Standard stars 
were selected from the catalog by \citet{hauck} and observed across each night.

The photometry was performed with DAOPHOT$\,${\footnotesize IV}/ALLFRAME \citep{ste87,ste94} 
and aperture photometry on the standards with ROMAFOT \citep{buonanno}.
Extinction coefficients were estimated from observations of standards 
at different air-mass values. Calibration curves from the two observing 
nights agreed quite well and we selected the best photometric night, 
July 6, as the reference night. 
The final calibrated catalog includes $\approx$80,000 stars with an 
accuracy of $\sigma_y \lesssim$ 0.1 and $\sigma_{v-y} \lesssim$ 0.2 mag 
at $y \approx$ 20 mag. The accuracy of the calibration is $\sim$ 0.02 mag for 
the $y,b,v$ bands and $\sim$ 0.05 mag for the $Ca,u$ bands.

\begin{figure*}[ht!]
\begin{center}
\label{fig1}
\includegraphics[height=0.5\textheight,width=0.95\textwidth]{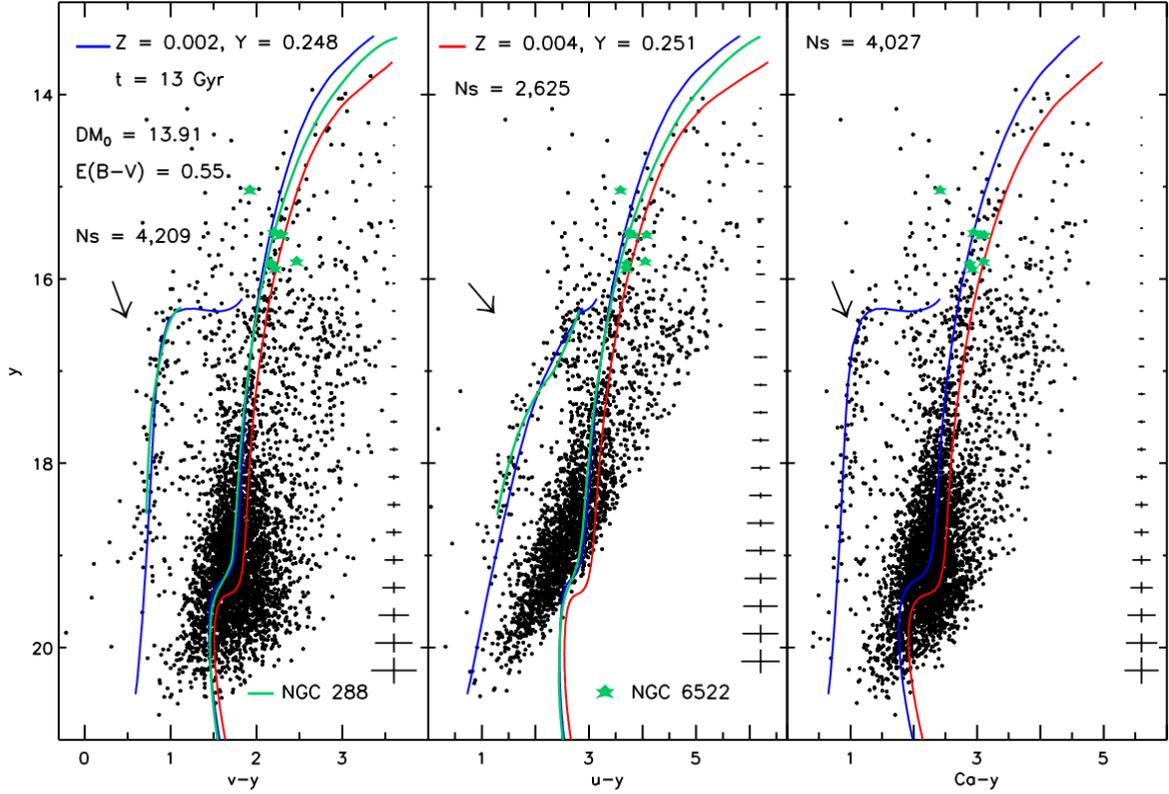}
\caption{$y$, \vmy, \umy, \camy\ CMDs of NGC~6522.
Stars were selected according to distances from the center 
(0.65\arcmin $\le$r$\le$ 1.65\arcmin) and photometric accuracy. 
Error bars display intrinsic errors in color and in magnitude, 
while the arrows show the reddening directions.
The red and the blue solid lines display two cluster isochrones at fixed 
age and for different chemical compositions and the predicted ZAHB for 
$Z=0.002$. The green solid lines show the ridge lines 
of NGC~288. The adopted true distance modulus and cluster reddening 
are labeled. The green stars mark the eight cluster RGs observed 
spectroscopically by \citet{bar09}.}
\end{center}
\end{figure*}

In this investigation we are focusing on NGC~6522, therefore
the final catalog was restricted by photometric accuracy and star position. 
Only stars with distances from the cluster center ($\alpha$ = 18:03:34, 
$\delta$ = -30:02:02) in the range 0.65\arcmin -- 1.65\arcmin~ were plotted 
in the $y$, \vmy\ (left panel), $y$, \umy\ (middle) and $y$, \camy\ (right) 
Color-Magnitude Diagrams (CMDs) of Fig.~1. The cluster center was excluded 
due to crowding, but stars up to about 1.5 the half-light radius 
($r_h = 1.0$\arcmin, \citealt{harris}) are selected. The entire photometric 
catalog will be presented in a forthcoming paper.

In order to validate current absolute calibration we compared 
cluster photometry with both theoretical predictions and \strom photometry 
of NGC~288 (CA07). We adopted this GC, because current spectroscopic measurements 
indicate an iron abundance (\feh = -1.32$\pm$0.02 dex) similar to 
NGC~6522 (\feh = -1.45$\pm$0.08 dex, \citealt{carretta09}).

We adopted a true distance modulus for NGC~6522 of  
$\mu_0=13.91$ and a mean reddening $E(B-V)=0.55$ \citep{bar98}.
The extinction coefficients for the \strom colors were estimated by applying 
the \citet{card89} 
reddening relation and $R_V = A_V/E(B-V) = 3.13$, 
according to the reddening dependence of $R_V$ on $E(B-V)$
\citep[see also Barbuy et al. 1998]{ols75}
We found:  
$E(b-y)= 0.69\times E(B-V)$, 
$E(v-y)= 1.31\times E(B-V)$, 
$E(u-y)= 1.82\times E(B-V)$ and
$E(Ca-y)= 1.46\times E(B-V)$.

The blue and the red solid lines in Fig.~1 show two cluster isochrones 
at fixed age ($t = 13$ Gyr) and different chemical compositions, 
namely $Z=0.002, Y=0.248$ and $Z=0.004, Y=0.251$, and the predicted 
Zero Age Horizontal Branch (ZAHB) for $Z=0.002$. Isochrones are from 
the BASTI data base and are based on $\alpha$-enhanced ($\afe=0.4$) 
evolutionary models \citep[hereafter PI06]{pietri06},
transformed into the observational plane using atmosphere models computed 
assuming $\alpha$-enhanced mixtures. Data plotted in 
Fig.~1 show that theory and observations, within the errors, agree 
quite well over the entire magnitude range. 

In particular, the two isochrones bracket the RGs with known metal 
abundance, i.e. $-0.96 \lesssim \mh \lesssim -0.66$ dex 
($-1.31 \lesssim \feh \lesssim -1.01$ dex). 
The small discrepancy between the ZAHB and the HB might be due to 
the effect of differential reddening, which also produces part of 
the RGB spread. The reddening vector is shown for each CMD in Fig.~1. 

The green solid lines show the ridge lines of NGC~288 along the RGB 
and the HB. To compare the two GCs we adopted a true distance modulus 
of $\mu_0=14.67$ \citep{ferraro} and a reddening of $E(B-V)=0.01$ (CA07). 
Unfortunately, $Ca$-band photometry for this cluster is not available. 

The RGB ridge line of NGC~288 is systematically bluer than RGs in NGC~6522, 
thus suggesting that the latter GC is slightly more metal-rich than the 
former one. This result agrees quite well with recent iron abundances for 
eight RGs (green stars) provided by \citet[hereafter BA09]{bar09}. 
The measurements are based on high-resolution spectra collected with 
FLAMES/GIRAFFE at the VLT (ESO) and give \feh = -1.0$\pm$0.2 dex on the
\citet{cargrat} metallicity scale.

The consistency between theoretical and empirical scenario is further supported 
by the evidence that the NGC~288 ridge lines agree quite well with more 
metal-poor ($Z=0.002$) evolutionary predictions. 

Independent Metallicity--Index--Color (MIC) relations are derived 
using cluster isochrones based on $\alpha$-enhanced evolutionary models (PI06). 
Theoretical predictions were transformed into the observational plane
by adopting bolometric corrections (BCs) and Color--Temperature 
Relations (CTRs) based on atmosphere models computed assuming the same heavy 
element abundances \citep[PI06,][]{CK06}. 
The Vega flux adopted is from \citet{CK94}\footnote{The complete set of BCs, CTRs 
and the Vega flux are available at http://wwwuser.oat.ts.astro.it/castelli}.
The metallicities adopted for the calibration are: 
Z=0.0001, 0.0003, 0.0006, 0.001, 0.002, 0.004, and 0.01. 
We neglected more metal-rich structures because the $hk$ index loses sensitivity 
in the metal-rich regime (\citealt[hereafter ATT98]{twa98}, and references therein).
The adopted Z values indicate the global abundance of heavy elements in the             
chemical mixture, with a solar metal abundance of ${(\zx)_\odot}=0.0245$.

To unredden the $hk$ index we adopted $E(hk)$ = -0.155$\times E(\bmy)$ 
\citep[hereafter AT91]{twa91}.
Together with the unreddened index, $hk_0$, we also 
derive independent MIC relations for the reddening-free parameter 
$[hk] = hk_0\, +\, 0.155\, \times (\bmy)$, to overcome deceptive uncertainties 
caused by differential reddening. 

Fig.~2 shows the seven isochrones plotted in the $[hk], \, \vmy$ plane,
covering the evolutionary phases from the base of the RGB to the tip. 
Note the nonlinearity of the $hk, \, \vmy$ relations for 
RGs and the decrease in sensitivity of the $hk$ index when moving from 
the metal-poor to the metal-rich regime ($\mh \sim$ -0.5 dex), 
as originally suggested by ATT98. 

To select the $hk_0$ and $[hk]$ values along the individual isochrones 
we follow the same approach adopted for the calibration of the 
$m_1$ index (CA07).  A multilinear regression fit was performed to 
estimate the coefficients of the MIC relations for the $hk_0$ and the 
$[hk]$ indices as a function of four Color Indices (CIs): $\bmy, \vmy, \camy, \umy$:

\begin{eqnarray*}
hk_{0} = \alpha\, + \beta\,\mh + \gamma\, CI_0 +
\delta\, (CI_0 \times hk_0) + \epsilon\, CI_0^2 + \\
\zeta\, hk_0^2 + \eta\, (CI_0^2 \times hk_0) + \theta\, (CI_0 \times hk_0^2) + 
\iota\, (CI_0^2 \times hk_0^2) + \\
\kappa\, (CI_0 \times \mh) + 
\lambda\, (hk_0 \times \mh)
\end{eqnarray*}  

where the symbols have their usual meaning. 
The adoption of eleven terms, compared to the four of the $m_1$
calibration, is due to the nonlinearity of the $hk_0$ vs $CI_0$
relations for RGs. The coefficients of the fits, together 
with their uncertainties, for the eight MIC relations, 
are listed in Table~1. 

\begin{figure}[!ht]
\begin{center}
\label{fig2}
\includegraphics[height=0.4\textheight,width=0.5\textwidth]{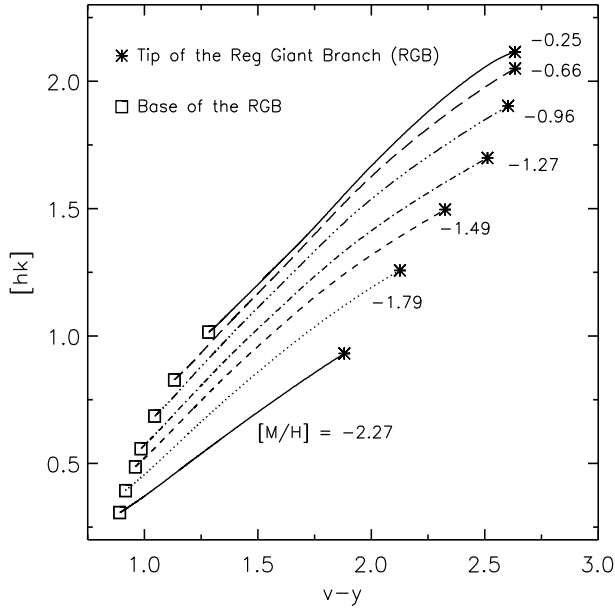}
\caption{$[hk]\,(\vmy)$ plane for isochrones at fixed cluster age ($t$=13 Gyr) 
and different chemical compositions ($\mh$, see labeled values).
The evolutionary phases range from the base of the RGB (squares) 
to the tip (asterisks). Evolutionary tracks were computed by assuming 
$\alpha-$enhanced chemical mixtures (PI06) and transformed 
by adopting atmosphere models with the same $\alpha-$enhancement.}
\end{center}
\end{figure}

\section{Validation of the new calibration of the $hk$ index}\label{validation}
In order to validate the new calibration of the $hk$ index we estimate 
the metallicity of field RGs for which \citet[hereafter ATT94]{twa94} and ATT98 collected both 
{\it Ca-uvby\/} photometry and high-resolution spectra. We end up with 
a sample of 96 RGs. For 28 of them we retrieve from the VO database
the $Ca$ and $Mg$ abundances of 
\citet[hereafter FU00]{fu00}, from which a proxy of the $\alpha$-enhancement 
is estimated either as $[\alpha/Fe] = [Ca/Fe]$ or as $[\alpha/Fe] = [(Ca+Mg)/Fe]$.
The metallicity range covered by our MIC relations 
is $-2.6<\feh<-0.6$ dex, but we select stars with $-2.7<\feh<-0.5$ dex to
account for uncertainties in spectroscopic abundances and in the 
metallicity scale \citep{kra03}. We end up with 85 RGs of which 
24 have $Ca$,$Mg$ abundance measurements.

We plot the difference between photometric and spectroscopic
metallicities for the 85 field RGs as a function of their spectroscopic
metal abundances in Fig.~3. Photometric abundances are estimated via
the $hk_0, \, \vmy$ (panels a,c) and the $hk_0, \, \camy$ relations (b,d). 
The global metallicity $\mh$ is estimated adopting the \citet{salaris} 
formula and either the $[Ca/Fe]$ (panels a,b) or the $[(Ca+Mg)/Fe]$ 
measurement (c,d) for the 24 RGs in common with FU00 (red dots), or 
a constant $\alpha$-enhancement for the remaining stars of 
$[\alpha/Fe]$=0.4 dex (RGs with $\feh \lesssim$ -0.8 dex)
and $[\alpha/Fe]$=0.15 dex (RGs with $\feh >$ -0.8 dex, black).
Data plotted in Fig.~3 show that the 
difference between spectroscopic measurements and photometric estimates is, 
on average, of the order of 0.1 dex when using the $hk_0, \,  \vmy$ relation 
($\sim$ -0.07$\pm$0.02 dex, $[\alpha/Fe] = [Ca/Fe]$; $\sim$ -0.09$\pm$0.02 dex, 
$[\alpha/Fe] = [(Ca+Mg)/Fe]$), or the $hk_0, \, \camy$ relation 
($\sim$ -0.09$\pm$0.02 dex; $\sim$ -0.11$\pm$0.02 dex). 
To overcome subtle uncertainties in the estimate of the mean difference 
we adopt the Biweight algorithm \citep{fabrizio}.  
The intrinsic dispersion of the different MIC relations is smaller than 
0.2 dex and caused either by photometric errors, or by reddening uncertainties, 
or by spectroscopic errors. The error bars in the bottom panel of Fig.~3 
display the mean error for the spectroscopic measurements 
($\sigma(\mh_{\hbox{\footnotesize spec}}) \sim$ 0.15 dex), estimated as 
the average of both the internal dispersion about the mean of $\feh$ measurements, 
the uncertainty due to the transformations into the standard metallicity scale 
(see column 8 in Table~2 and column 7 in Table~4 of ATT98), and the internal 
uncertainties of the $[Ca/Fe]$ and $[Mg/Fe]$ measurements by FU00.
The $CH$-strong stars (HD $\,$55496, HD$\,$135148, BD-01~2582, BD+04~2466, 
CD-62~1346, diamonds) do not show, in contrast with the metallicity based on the
$m_1$ MIC relations (see CA07), any peculiar discrepancy between photometric 
and spectroscopic metallicities. 
On the other hand, star HD$\,$84903 (asterisk), 
which might be affected by weak chromospheric emission in the core of the $Ca II$ $K$ 
line (ATT98), showed a large discrepancy ($\Delta \mh \approx$ -0.4 dex)
when using the $m_1$ MIC relations. We now adopt for HD$\,$84903 \feh = -2.6 dex, 
estimated accounting non-LTE effects by \citet{thevenin}. 
The difference between photometric and spectroscopic abundances 
is now inside current uncertainties ($\Delta \mh \approx$ -0.2 dex).
The star HD$\,$44007 (cross) was already discussed in ATT98, since the reddening 
correction is still uncertain. 
Photometric metallicities for two metal-rich stars (HD$\,$35179, 
HD$\,$7595) with $\mh >$ -0.60 dex, are 
systematically more metal-poor by $\sim$ 0.5 dex than spectroscopic 
measurements. Such a discrepancy might be due either to reddening 
uncertainties, or to the reduced sensitivity of the $hk$ index in 
the metal-rich regime.

\begin{figure}[!ht]
\begin{center}
\label{fig3}
\includegraphics[height=0.7\textheight,width=0.55\textwidth]{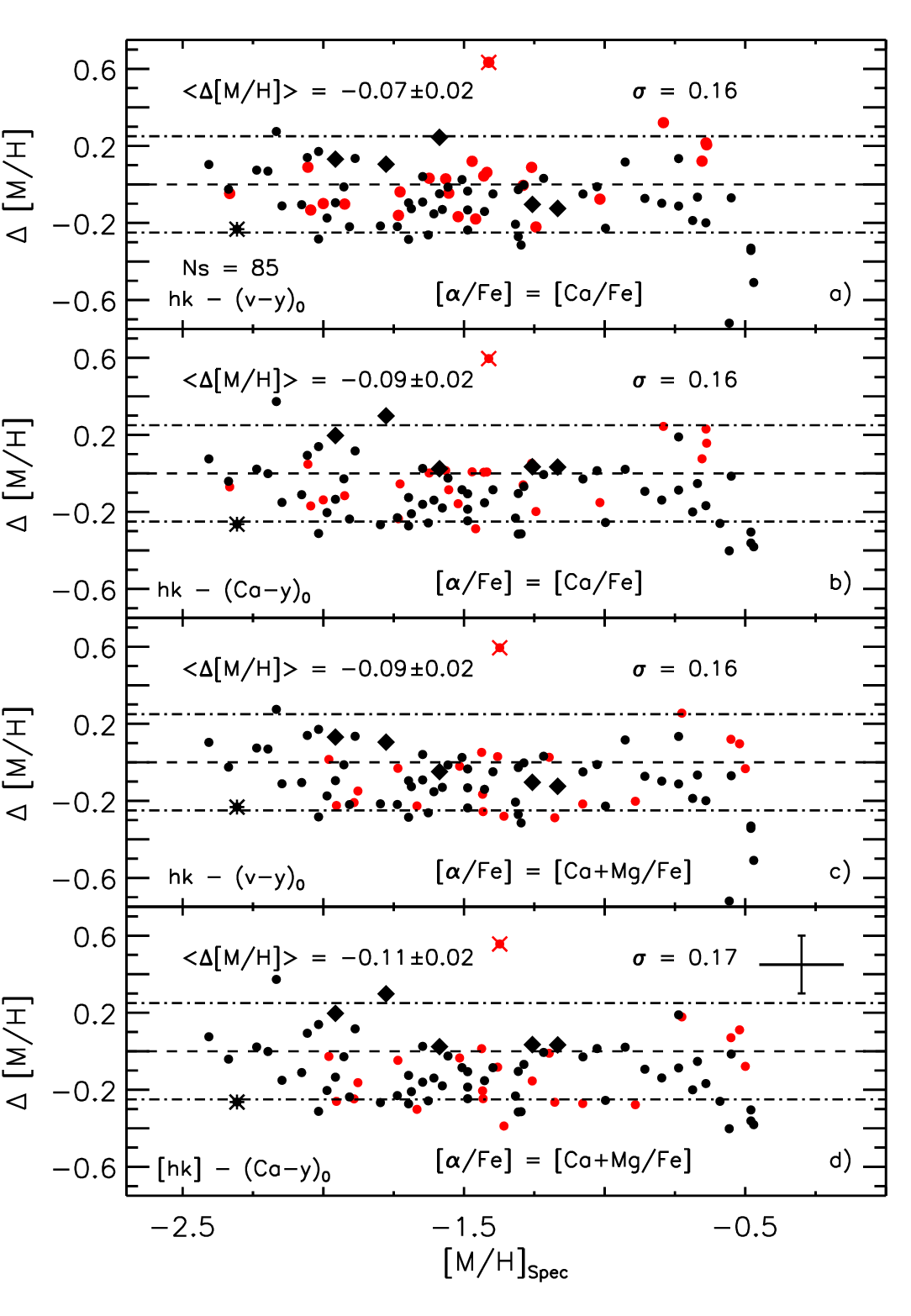} 
\caption{Difference between photometric and spectroscopic metallicities,
$\Delta \mh = (\mh_{\hbox{\footnotesize phot}} - \mh_{\hbox{\footnotesize spec}})$, 
plotted versus $\mh_{\hbox{\footnotesize spec}}$ for 85 field RGs by ATT94 
and ATT98. Panels a) and c) display photometric metallicities based on the 
$hk_0, \, \vmy$ relation, while panels b) and d) on the $hk_0,  \, \camy$ relation. 
For 24 RGs the $\alpha$-enhancement is estimated using either $Ca$ (panels a),b)) 
or $Ca$ and $Mg$ measurements (panels c),d)), while for the other objects a 
constant $\alpha$-enhancement is assumed (see text for more details). 
The diamonds mark the $CH$-strong stars, the asterisk star HD$\,$84903, 
while the cross the star with an uncertain reddening (HD$\,$44007). 
The error bars in the bottom panel 
display the mean error for the spectroscopic measurements.}
\end{center}
\end{figure}

We were not able to validate current calibrations with the spectroscopic
abundances of RGs in Baade's Window field \citep{zo08},  
since almost all of them are more metal-rich than \feh = -0.5 dex, therefore 
outside the metallicity range covered by the new MIC relations.

\begin{figure}[!ht]
\begin{center}
\label{fig4}
\includegraphics[height=0.6\textheight,width=0.5\textwidth]{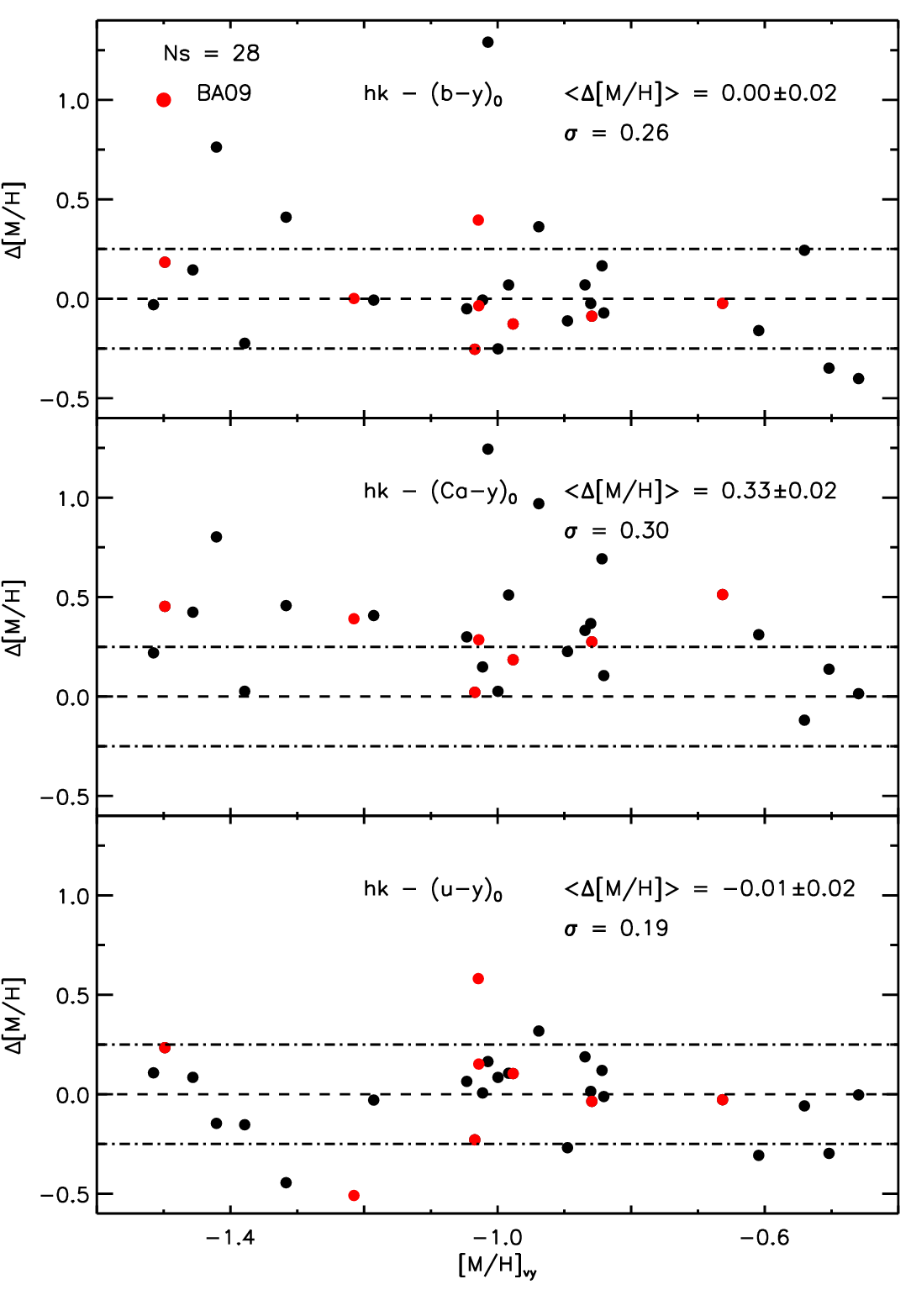} 
\caption{Difference between photometric metallicities estimated 
adopting different MIC relations, $hk, \, \bmy (top), $hk, \, \camy (middle) 
and $hk,\ , \umy$ (bottom) plotted versus the metallicity estimated with the
$hk, \, \vmy$ relation for 28 candidate cluster RGs. 
The eight RGs with spectroscopic measurements are marked in red.
}
\end{center}
\end{figure}

To further validate current calibrations, we apply the MIC relations 
to estimate the metallicity distribution of RGs in NGC~6522. 
The catalog is selected in star position as described 
in \S 1, in magnitude ($y <$ 18.0 mag), in photometric accuracy 
($\sigma (v,b,y) <$ 0.03 and $\sigma (Ca) <$ 0.02 mag), and 
in surface gravity ($[c]=c_1-0.2\times(b-y) <$ 0.35 mag), ending up with 51 RGs.
We downloaded the catalog of proper motions across the Galactic bar  
by \citet{sumi}
from the VO database, for a region of 15\arcmin$\times$15\arcmin~ 
across the cluster center. The match with our catalog gives $\sim13,200$ stars in 
common. The accuracy of the proper motions is $\sim$ 1 mas/yr, while the accuracy 
of the positional match is $\lesssim$ 1\arcsec.
We further select our sample with $-2 < pm(\alpha) < 6$ and 
$-2 < pm(\delta) < 2$ mas/yr, following the classification made by Sumi et al. 
of RG, red-clump and disc stars (see their Fig.~8).
Since the proper motions of field stars partly overlap with those of 
cluster stars, we cannot exclude contamination in the final sample 
of 28 candidate cluster RGs.  

Fig.~4 shows the difference between photometric metallicities of the 28 
candidate RGs estimated adopting the $hk_0, \, \vmy$ MIC relation and the
relations based on $hk_0$ and the $\bmy, \, \camy, \, \umy$ colors 
plotted versus metallicities estimated with $hk_0, \, \vmy$.  

By using the eight RGs (red dots) in common with the spectroscopic sample (BA09), 
we find that the difference (Biweigth mean) between photometric and 
spectroscopic metallicity\footnote{The $\alpha$-element abundance of the 
spectroscopic RGs is estimated as $[\alpha/Fe]=[(Ca+Si+Mg)/Fe]$, 
while their \feh~ is transformed in the Zinn \& West (1984) 
scale using the Carretta \& Gratton (1997) relation.}
is minimal not only for the $hk_0, \, \vmy$ 
($<\Delta(\mh_{\hbox{\footnotesize phot}} - \mh_{\hbox{\footnotesize spec}})>\, 
\simeq -0.02\pm0.01$, $\sigma$=0.26 dex), 
but also for the $hk_0, \, \bmy$ ($\simeq -0.05\pm0.01$, $\sigma$=0.38 dex)  
and the $hk_0, \, \umy$ ($\simeq -0.03\pm0.01$, $\sigma$=0.49 dex) MIC relation.  
The difference becomes larger when using the 
$hk_0, \, \camy$ ($\simeq 0.22\pm0.01$, $\sigma$=0.45 dex) relation.  
The difference is marginally larger when using the MIC relations based on the  
reddening free index. In particular, it is 
$\simeq -0.12\pm0.01$, $\sigma$=0.67 for the $[hk],\ \camy$ and  
$\simeq -0.17\pm0.01$, $\sigma$=0.43 for the $[hk],\ \umy$ relations. 
The difference between unreddened and reddening free MIC relations is mainly 
due to the sum in quadrature of the intrinsic photometric error. 
The evidence that the MIC relations using the $Ca$-band both in the 
metallicity index and in the color index show either the largest difference 
or the largest dispersion suggest that the discrepancy might be intrinsic.  
Data plotted in Fig.~4 further support this finding. Indeed, metallicity 
estimates based on the $hk_0, \, \bmy$ and on the $hk_0, \, \umy$ relations   
agree reasonably well with those based on the $hk_0, \, \vmy$ relation 
($<\Delta(\mh_{\hbox{\footnotesize by, uy}} - \mh_{\hbox{\footnotesize vy}})>\, 
\simeq 0.0\pm0.02$ and $\simeq -0.01\pm0.02$ with $\sigma$ = 0.26 and 0.19 dex, 
respectively.)
On the other hand, the metallicity estimates based on the $hk_0, \, \camy$ 
relation are on average $\approx$0.3 dex more metal-rich than those 
based on the $hk_0, \, \vmy$ relation.    
The difference might be due to the fact that the $hk_0, \, \camy$ relation is more 
sensitive to the $Ca$ abundance than the other relations. 
Moreover, \citet{twa95} suggested that the $hk$ index 
of RGs in M22 might be affected by a continuous absorption 
in the wavelength range between $3,900$ and $4,100$ \AA~ (see also \citealt{bondneff}).
We plan to provide a more quantitative analysis of this effect 
in a forthcoming paper.
Another possible culprit might be the chromospheric activity, 
since this phenomenon causes an emission in the core of the $Ca II$ $K$ 
line (\citealt{smith92, dupree, lee09a}). 
Unfortunately, the high-resolution spectra collected by BA09 
do not cover the wavelength region of $Ca II$ $H$ $K$ lines.

The photometric metallicity distributions show a well-defined main peak 
around $\mh \sim -0.95$ dex and two shoulders at $\mh \sim$ -1.4 and -0.5 dex. 
These features agree quite well with the spectroscopic metallicity distribution,
which shows two peaks at $\mh \sim -1.1$ and $\mh \sim -0.85$ dex, 
with the latter one including $\sim$ 10\% of the stars.
The difference between the main and the secondary peak might be due to 
an enhancement either in $Ca$ or in other heavy elements.
We cannot reach a firm conclusion concerning the few more metal-poor 
($\mh \lesssim$ -1.4 dex) outliers, since the standard deviations of the 
MIC relations range from $\sim$ 0.25 to 0.5 dex. They might be either 
field bulge stars or objects affected by differential reddening. 

\citet{lee09a,lee09b} found a double peaked distribution when applying 
their $hk_0, \, \bmy$ metallicity relation to stars in NGC~1851.
This is a peculiar GGC, with a split along the sub-giant and the 
RG branches, that might be due to the presence of two stellar populations with 
different $CNO$ abundance (CA07; \citealt{cassisi}). The secondary
peak in the metallicity distribution includes $\sim$ 18\% of the RGs 
\citep{lee09a}, including three $Ca$-enhanced stars \citep{yong}
and three $CN$-strong stars \citep{hesser} with enhanced abundances of 
$Ba$ and $Sr$. 

The above results indicate that possible differences in cluster RG
colors including the $Ca$-band should be cautiously treated, since they
might be caused either by changes in $Ca$ and/or by 
other heavy elements or by molecular bands.

We show that the \strom $hk$ index is a good diagnostic to estimate 
the global metal abundance of field and cluster RGs, and it can be also
adopted to detect stars affected by $Ca$ enhancement.
Moreover, the $hk$ index is more sensitive than the $m_1$ index in the 
metal-poor regime, and is less affected by $CN, CH$ peculiarities. 
The current MIC relations have been validated by adopting RGs in 
NGC~6522 and field RGs with known spectroscopic abundances, and provide 
metallicities with an accuracy better than 0.2 dex.
The application of the new MIC relations appear very promising not only 
for RGs in halo GGCs, but also to pin point metal-poor stars in the 
Galactic halo.  

\acknowledgements
It is a real pleasure to thank F. Castelli for sending us
bolometric corrections and color indices for \strom bands.
MZ is partly supported by Proyecto FONDECYT Regular 1110393, the
FONDAP Center for Astrophysics 1510003, the BASAL CATA PFB-06, the
Milky Way Millennium Nucleus from ICM grant P07-021-F, and by
Proyecto Conicyt Anillo ACT-86.


\bibliographystyle{apj}

\begin{deluxetable}{lccccccccccc}
\tablewidth{0pt}
\tabletypesize{\scriptsize}
\tablecaption{Multilinear regression coefficients for the \strom  
metallicity index: $hk_{0} = \alpha\, + \beta\,\feh + \gamma\, CI_0 +
\delta\, (CI_0 \times hk_0) + \epsilon\, CI_0^2 + 
\zeta\, hk_0^2 + \eta\, (CI_0^2 \times hk_0) + \theta\, (CI_0 \times hk_0^2) + 
\iota\, (CI_0^2 \times hk_0^2) + 
\kappa\, (CI_0 \times \feh) + 
\lambda\, (hk_0 \times \feh)$.\label{tbl-1}}
\tablehead{
\colhead{Relation}&
\colhead{$\alpha$}&
\colhead{$\beta$}&
\colhead{$\gamma$}&
\colhead{$\delta$}&
\colhead{$\epsilon$}&
\colhead{$\zeta$}&
\colhead{$\eta$}&
\colhead{$\theta$}&
\colhead{$\iota$}&
\colhead{$\kappa$}&
\colhead{$\lambda$}\\
\colhead{(1)}&
\colhead{(2)}&
\colhead{(3)}&
\colhead{(4)}&
\colhead{(5)}&
\colhead{(6)}&
\colhead{(7)}&
\colhead{(8)}&
\colhead{(9)}&
\colhead{(10)}&
\colhead{(11)}&
\colhead{(12)}
}
\startdata
\multicolumn{12}{}{}  \\  
$hk_0, (\bmy)_0$ & 0.125 & 0.116 & 0.756 & 0.860 & -0.556 & 0.684 & -0.282 & -1.057 & 0.495 & -0.019 & -0.081 \\    
Error            & 0.010 & 0.009 & 0.172 & 0.297 &  0.142 & 0.047 &  0.144 &  0.045 & 0.032 &  0.027 & 0.012 \\    
$hk_0, (\vmy)_0$ & -0.009 & 0.096 & 0.615 & -0.072 & -0.107 & 0.735 & 0.040 & -0.299 & 0.041 & 0.047 & -0.133 \\    
Error            &  0.010 & 0.007 & 0.056 &  0.091 &  0.026 & 0.056 & 0.021 &  0.014 & 0.004 & 0.013 &  0.015 \\
$hk_0, (\camy)_0$ & -0.028 & 0.051 & 0.425 & -0.094 & -0.034 & 0.720 & 0.018 & -0.176 & 0.016 & 0.050 & -0.131 \\    
Error             & 0.010  & 0.010 & 0.035 &  0.140 &  0.032 & 0.155 & 0.010 &  0.018 & 0.002 & 0.017 &  0.029 \\
$hk_0, (\umy)_0$ & 0.131 & 0.097 & 0.183 & 0.125 & -0.018 & 0.616 & -0.011 & -0.171 & 0.017 & 0.012 & -0.099 \\    
Error            & 0.010 & 0.008 & 0.031 & 0.054 &  0.008 & 0.062 &  0.007 &  0.008 & 0.001 & 0.008 &  0.016 \\
$[hk], (\bmy)_0$ & 0.162 & 0.105 & 0.657 & 1.030 & -0.375 & 0.547 & -0.493 & -0.864 & 0.433 & 0.011 & -0.070 \\    
Error            & 0.010 & 0.009 & 0.181 & 0.191 &  0.152 & 0.039 &  0.133 &  0.035 & 0.029 & 0.027 &  0.011 \\
$[hk], (\vmy)_0$ & 0.022 & 0.086 & 0.597 & -0.057 & -0.060 & 0.644 & 0.005 & -0.247 & 0.039 & 0.064 & -0.132 \\    
Error            & 0.010 & 0.006 & 0.058 &  0.091 &  0.031 & 0.051 & 0.019 &  0.012 & 0.004 & 0.013 &  0.014 \\
$[hk], (\camy)_0$ & -0.021 & 0.049 & 0.462 & -0.147 & -0.030 & 0.703 & 0.023 & -0.157 & 0.013 & 0.053 & -0.126 \\    
Error             & 0.010  & 0.010 & 0.038 &  0.172 &  0.043 & 0.174 & 0.011 &  0.017 & 0.002 & 0.019 &  0.031 \\
$[hk], (\umy)_0$ & 0.127 & 0.083 & 0.206 & 0.090 & -0.009 & 0.580 & -0.013 & -0.149 & 0.015 & 0.025 & -0.106 \\    
Error            & 0.010 & 0.008 & 0.029 & 0.056 &  0.009 & 0.059 &  0.005 &  0.007 & 0.001 & 0.009 &  0.017 \\
\enddata
\end{deluxetable}				   


\end{document}